\documentclass[dvips]{article}
\usepackage{icrctc07}

\title{Future plan for observation of cosmic gamma rays in the 100 TeV energy region \\
with the Tibet air shower array : simulation and sensitivity
}
\shorttitle{Future plan of the Tibet experiment}
\authors{
The Tibet AS$\gamma$ Collaboration\\
M.~Amenomori$^{1}$, X.~J.~Bi$^{2}$, D.~Chen$^{3}$, S.~W.~Cui$^{4}$,
Danzengluobu$^{5}$, L.~K.~Ding$^{2}$, X.~H.~Ding$^{5}$, C.~Fan$^{6}$,
C.~F.~Feng$^{6}$, Zhaoyang Feng$^{2}$, Z.~Y.~Feng$^{7}$,
X.~Y.~Gao$^{8}$, Q.~X.~Geng$^{8}$, H.~W.~Guo$^{5}$, H.~H.~He$^{2}$,
M.~He$^{6}$, K.~Hibino$^{9}$, N.~Hotta$^{10}$, Haibing~Hu$^{5}$,
H.~B.~Hu$^{2}$, J.~Huang$^{11}$, Q.~Huang$^{7}$, H.~Y.~Jia$^{7}$,
F.~Kajino$^{12}$, K.~Kasahara$^{13}$, Y.~Katayose$^{3}$,
C.~Kato$^{14}$, K.~Kawata$^{11}$, Labaciren$^{5}$, G.~M.~Le$^{15}$,
A.~F.~Li$^{6}$, J.~Y.~Li$^{6}$, Y.-Q.~Lou$^{16}$, H.~Lu$^{2}$,
S.~L.~Lu$^{2}$, X.~R.~Meng$^{5}$, K.~Mizutani$^{13,17}$, J.~Mu$^{8}$,
K.~Munakata$^{14}$, A.~Nagai$^{18}$, H.~Nanjo$^{1}$,
M.~Nishizawa$^{19}$, M.~Ohnishi$^{11}$, I.~Ohta$^{20}$,
H. Onuma$^{17}$, T.~Ouchi$^{9}$, S.~Ozawa$^{11}$, J.~R.~Ren$^{2}$,
T.~Saito$^{21}$, T.~Y.~Saito$^{22}$, M.~Sakata$^{12}$,
T.~K.~Sako$^{11}$, M.~Shibata$^{3}$, A.~Shiomi$^{9,11}$,
T.~Shirai$^{9}$, H.~Sugimoto$^{23}$, M.~Takita$^{11}$,
Y.~H.~Tan$^{2}$, N.~Tateyama$^{9}$, S.~Torii$^{13}$,
H.~Tsuchiya$^{24}$, S.~Udo$^{11}$, B.~Wang$^{8}$, H.~Wang$^{2}$,
X.~Wang$^{11}$, Y.~Wang$^{2}$, Y.~G.~Wang$^{6}$, H.~R.~Wu$^{2}$,
L.~Xue$^{6}$, Y.~Yamamoto$^{12}$, C.~T.~Yan$^{11}$, X.~C.~Yang$^{8}$,
S.~Yasue$^{25}$, Z.~H.~Ye$^{15}$, G.~C.~Yu$^{7}$, A.~F.~Yuan$^{5}$,
T.~Yuda$^{9}$, H.~M.~Zhang$^{2}$, J.~L.~Zhang$^{2}$,
N.~J.~Zhang$^{6}$, X.~Y.~Zhang$^{6}$, Y.~Zhang$^{2}$, Yi~Zhang$^{2}$,
Zhaxisangzhu$^{5}$ and X.~X.~Zhou$^{7}$
}

\shortauthors{M.~Amenomori {\it et al.}}
\afiliations{\noindent
$^{1}$Department of Physics, Hirosaki University, Hirosaki 036-8561, Japan.
$^{2}$Key Laboratory of Particle Astrophysics, Institute of High Energy Physics, Chinese Academy of Sciences, Beijing 100049, China.
$^{3}$Faculty of Engineering, Yokohama National University, Yokohama 240-8501, Japan.
$^{4}$Department of Physics, Hebei Normal University, Shijiazhuang 050016, China.
$^{5}$Department of Mathematics and Physics, Tibet University, Lhasa 850000, China.
$^{6}$Department of Physics, Shandong University, Jinan 250100, China.
$^{7}$Institute of Modern Physics, SouthWest Jiaotong University, Chengdu 610031, China.
$^{8}$Department of Physics, Yunnan University, Kunming 650091, China.
$^{9}$Faculty of Engineering, Kanagawa University, Yokohama 221-8686, Japan.
$^{10}$Faculty of Education, Utsunomiya University, Utsunomiya 321-8505, Japan.
$^{11}$Institute for Cosmic Ray Research, University of Tokyo, Kashiwa 277-8582, Japan.
$^{12}$Department of Physics, Konan University, Kobe 658-8501, Japan.
$^{13}$Research Institute for Science and Engineering, Waseda University, Tokyo 169-8555, Japan.
$^{14}$Department of Physics, Shinshu University, Matsumoto 390-8621, Japan.
$^{15}$Center of Space Science and Application Research, Chinese Academy of Sciences, Beijing 100080, China.
$^{16}$Physics Department and Tsinghua Center for Astrophysics, Tsinghua University, Beijing 100084, China.
$^{17}$Department of Physics, Saitama University, Saitama 338-8570, Japan.
$^{18}$Advanced Media Network Center, Utsunomiya University, Utsunomiya 321-8585, Japan.
$^{19}$National Institute of Informatics, Tokyo 101-8430, Japan.
$^{20}$Tochigi Study Center, University of the Air, Utsunomiya 321-0943, Japan.
$^{21}$Tokyo Metropolitan College of Industrial Technology, Tokyo 116-8523, Japan.
$^{22}$Max-Planck-Institut f\"ur Physik, M\"unchen D-80805, Deutschland.
$^{23}$Shonan Institute of Technology, Fujisawa 251-8511, Japan.
$^{24}$RIKEN, Wako 351-0198, Japan.
$^{25}$School of General Education, Shinshu University, Matsumoto 390-8621, Japan.
}

\email{tsako@icrr.u-tokyo.ac.jp}

\abstract{
The Tibet air shower array, which has an effective area of 37,000 square meters and 
is located at 4300 m in altitude, has been observing air showers 
induced by cosmic rays with energies above a few TeV.
We have a plan to add a large muon detector array to it for the purpose of increasing its sensitivity 
to cosmic gamma rays in the 100 TeV energy region by discriminating them from cosmic-ray hadrons.
We have deduced the attainable sensitivity of the muon detector array using our Monte Carlo simulation.
We report here on the detailed procedure of our Monte Carlo simulation.
}

\begin{document}
\maketitle

\section{Introduction}
Supernova remnants (SNRs) are the best candidates for acceleration of hadronic cosmic rays 
up to the knee in the cosmic-ray energy spectrum at $\sim10^{15}$eV. Consequently, gamma rays in the 100 TeV 
region (10 - 1000 TeV) originating in $\pi^0$ decay following inelastic collisions 
between accelerated charged cosmic rays and the ambient medium are naturally expected.
Since the expected flux of $\pi^0$ decay gamma rays is low, an apparatus with a high duty-cycle and a large
field of view, e.g. an air shower array, is suitable for this purpose.  
Air shower arrays thus far, however, have not been sensitive enough for the detection of such gamma rays.
This is because their angular resolution and/or their power of discriminating gamma-ray induced air showers 
from hadron-induced ones were insufficient.

The Tibet air shower (AS) array has succeeded in detecting celestial TeV gamma rays \cite{allsky}
and is operating with effective area of 37,000 m$^2$.
At the 100 TeV energy, its angular resolution and energy resolution are estimated to be 0.2$^{\circ}$ 
and 40\%, respectively.
We are now planning to build a large water Cherenkov muon detector (MD) array 
in the underground of the AS array for the purpose of gamma-hadron discrimination.
This future plan would be able to open the 100 TeV energy window in gamma-ray astronomy.
\section{Muon Detector Design}
\begin{figure}
  \begin{center}
    \includegraphics [bb=0 80 590 620, width=0.48\textwidth]{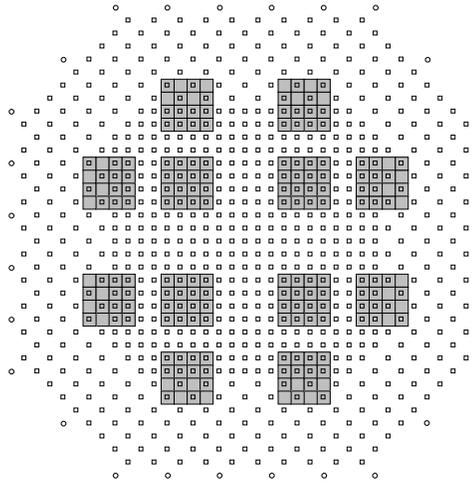}
  \end{center}
  \caption{Schematic view of the Tibet AS+MD array. Open squares and open circles represent 
the surface scintillation detectors that compose the Tibet AS array.
Note that the AS array drawn here is upgraded from the current version so that its effective area 
becomes 50,000 m$^2$ by modifying the configuration of the scintillation detectors. 
Filled squares show the proposed Tibet MD array 2.5 m underground.}\label{config}
\end{figure}

The currently proposed configuration of the MD array is shown in Figure \ref{config}.
It is composed of 12 pools, each of which consists of 16 cells.
Each cell is a waterproof concrete tank which is 7.2 m wide $\times$ 7.2 m long $\times$ 1.5 m deep 
in size. 
Two 20 inch-in-diameter photomultiplier tubes (PMTs, Hamamatsu R3600) are put on its ceiling, 
facing downwards. 
Its inside is painted with white epoxy resin to waterproof and to efficiently reflect 
water Cherenkov light, which is then collected with the PMTs.
The MD array is set up 2.5 m underground (2.0 m soil + 0.5 m concrete ceiling, $\sim$19 radiation lengths)
in order to detect the penetrating muon component of air showers, suppressing the electromagnetic one.
Its total effective area amounts to 9,950 m$^2$ for muon detection with the energy threshold of 
approximately 1 GeV.
The advantages of using the water Cherenkov type detector are high cost performance and its capability
to exclude the influence of the environmental background radioactivity thanks to the Cherenkov threshold.

\section{Simulation}
The air shower events induced by primary cosmic rays and gamma rays were generated 
in the energy range from 0.3 TeV to 10 PeV and within zenith angle less than 60$^{\circ}$ 
along the Crab's orbit using the Corsika Ver.6.204 code \cite{corsika}. 
We used QGSJET01c for the hadronic interaction model and adopted a chemical composition model
\cite{HD4} based on direct observational data for the primary cosmic rays.
We assumed a differential energy spectrum $E^{-2.6}$ for the primary gamma rays.
Air shower events were uniformly thrown within 300 m from the array center.
This radius is sufficient to collect all the air shower events which actually trigger the AS array.
The response of the AS array was simulated using the Epics uv8.00 code \cite{Epics}, which had been already 
established. From the simulation, we deduced the air shower direction, core position and the sum of the 
number of particles per m$^2$ detected in each scintillation counter ($\sum\rho$) etc.
Note that in this simulation we used an upgraded version of the AS array, the effective area of which was
increased from 37,000 m$^2$ to 50,000 m$^2$ by modifying the configuration of the 
current scintillation detectors.
The total number of the scintillation detectors were not changed.
\begin{figure}[t]
  \begin{center}
    \includegraphics [width=0.48\textwidth]{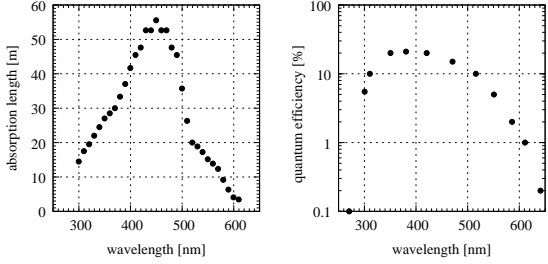}
  \end{center}
  \caption{The light attenuation length (left) and the quantum efficiency of the PMTs (right). 
  Both are dependent on wavelength of light.}
\label{fig1}
\end{figure}

We selected air shower events based on several conditions, i.e. software trigger condition of 
any fourfold coincidence in the FT counters recording more than 1.25 particle in charge, 
air shower core position located in the array and
the residual error of direction reconstruction less than 1.0 m.

The secondary particles of the surviving air shower events were subsequently fed into the simulation of 
the soil absorber, then the MD array. Their responses were simulated based on GEANT4 8.0 code \cite{GEANT}.
We assumed that the soil was a mixture of 70\% SiO$_2$, 20\% Al$_2$O$_3$ and 10\% CaO 
with density 2.0 g/cm$^2$ and thickness 2.0 m. 
For the MD array, its detailed structure made of concrete (2.3 g/cm$^3$, 100\% SiO$_2$) 
was taken into account and the ceiling thickness of each cell was set to be 0.5 m.
The reflectance on the inner surface of the cells was assumed to be 70\% with isotropic reflection.
The light attenuation length and the quantum efficiency of the 20 inch PMTs used in the simulation 
are shown in Figure \ref{fig1}.
After simulating Cherenkov radiation, propagation of Cherenkov photons in water and the response of the PMTs,
the number of the collected photoelectrons ($N_{\mbox{\scriptsize PE}}$) was obtained for each muon detector.
$N_{\mbox{\scriptsize PE}}$ resolution for one vertical penetrating muon is 
estimated to be 34 PEs$^{+180\%}_{-18\%}$ from the simulation.

\section{Results and discussions}
\begin{figure}[t]
  \begin{center}
    \includegraphics [width=0.48\textwidth]{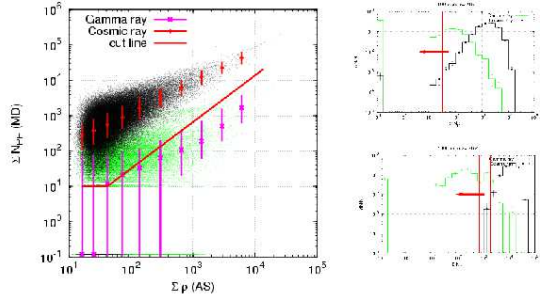}
  \end{center}
  \caption{Left : Distribution of $N_{\mbox{\scriptsize PE}}$ as a function of $\sum\rho$.
Green and black dots correspond to gamma-induced and hadron-induced air shower events, respectively.
Each closed circle with an error bar represents the 20\%, 50\%, 80\% of the median distribution 
in each $\sum\rho$ bin. The solid line shows the optimized cut to suppress hadron-induced events.
Air showers accompanied by no PEs are plotted at $N_{\mbox{\scriptsize PE}}$ = 1.2.
Upper right : $N_{\mbox{\scriptsize PE}}$ distribution in the 10 TeV energy band 
($100 \le \sum\rho < 215$). Lower right : that in the 100 TeV energy band ($1000 \le \sum\rho < 2154$).
}
\label{scat}
\end{figure}

Figure \ref{scat} shows the distribution of $\sum N_{\mbox{\scriptsize PE}}$ as a function of $\sum\rho$.
$\sum N_{\mbox{\scriptsize PE}}$ denotes the sum of $N_{\mbox{\scriptsize PE}}$ for muon detectors fired 
with $N_{\mbox{\scriptsize PE}} > 10$. $\sum\rho = 1000$ corresponds to approximately 100 TeV 
primary gamma-ray energy. 
$\sum N_{\mbox{\scriptsize PE}}$ allows for the influence of accidental muons, which impinges each
muon detector at the rate of nearly 1.6 kHz.
In selecting muon-poor events, we set the optimized cut condition as shown in Figure \ref{scat}.

\begin{figure}
  \begin{center}
    \includegraphics [bb=15 25 505 485, width=0.48\textwidth]{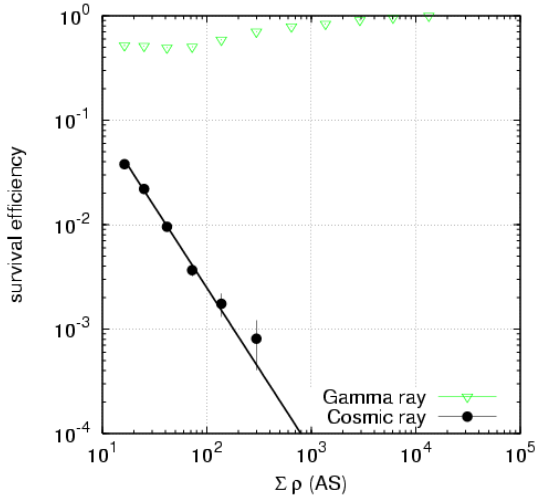}
  \end{center}
  \caption{ The survival efficiency after the cut. Green and black circles represent gamma-induced
and hadron-induced events, respectively.
}
\label{survival}
\end{figure}

\begin{figure}
  \begin{center}
    \includegraphics [bb=50 70 555 530,width=0.48\textwidth]{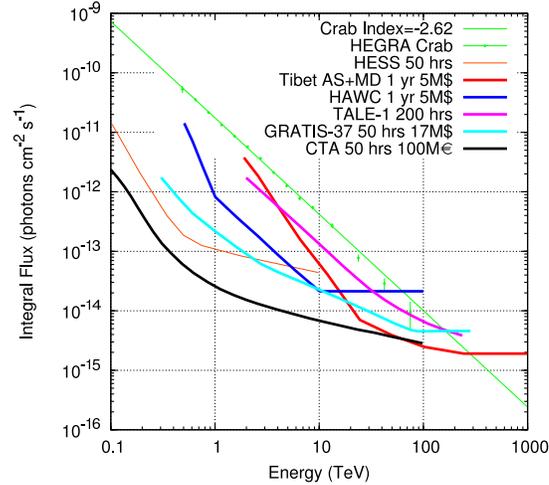}
  \end{center}
  \caption{ The attainable integral flux sensitivity of the Tibet AS+MD array 
to a point-like gamma-ray source, together with the sensitivity of HESS and some other future plans.
}
\label{sens}
\end{figure}

Figure \ref{survival} shows the survival efficiency of the air shower events after the cut, obtained from
the simulation.
Around $\sum\rho = 1000$, the number of hadron-induced events are suppressed down to 0.01\% or less, while 
gamma-induced events are retained by more than 83\%.
Finally, Figure \ref{sens} shows the attainable integral flux sensitivity of the Tibet AS+MD array to 
a point-like gamma-ray source.
Its 5$\sigma$ sensitivity in one calendar year will reach 7\% and $\sim$20\% Crab 
above 20 and 100 TeV respectively, and surpass the existing IACTs above 20 TeV.
Furthermore, it may surpass the next generation IACTs above 40 TeV.
A further discussion is conducted in \cite{myoral}.
The Tibet AS+MD array will contribute to a deeper understanding of the origin 
and the acceleration mechanism of cosmic rays in cooperation with other experiments.

\section{Acknowledgements}
The collaborative experiment of the Tibet Air Shower Arrays has been
performed under the auspices of the Ministry of Science and Technology
of China and the Ministry of Foreign Affairs of Japan. This work was
supported in part by Grants-in-Aid for Scientific Research on Priority
Areas (712) (MEXT), by the Japan Society for the Promotion of Science,
by the National Natural Science Foundation of China, and by the
Chinese Academy of Sciences.


\begin{thebibliography}{20}
\bibitem{allsky} M. Amenomori et al., 2005, ApJ 633, 1005.
\bibitem{kawata} M. Amenomori et al., astro-ph/0611030.
\bibitem{corsika} D. Heck et al., Forschungszentrum Karlsruhe Report 
                  No. FZKA 6019, 1998.
\bibitem{HD4} M. Amenomori et al., 2006, Advances in Space Research 37, 1932.
\bibitem{Epics} {\small EPICS}, http://cosmos.n.kanagawa-u.ac.jp/EPICSHome/, 2003.
\bibitem{GEANT} Agostinelli S. et al., 2003, NIM A 506, 250.
\bibitem{myoral} M. Amenomori et al., to appear in the proceedings of 30th ICRC : a talk entitled by 
``Future plan for observation of cosmic gamma rays in the 100 TeV energy region 
with the Tibet air shower array : physics goal and overview''.
\end{thebibliography}

\bibliographystyle{plain}

\end{document}